\documentclass{aa}

\usepackage{amssymb}

\usepackage{graphicx}

\begin{document}

\title{Cosmic ray acceleration by spiral shocks in the galactic wind}
\author{H.J. V\"olk, \inst{1} \and V.N. Zirakashvili \inst{2}}
\date{Received ; accepted }


\subtitle{}

\offprints{V.N. Zirakashvili}

\institute{Max-Planck-Institut f\"{u}r\ Kernphysik, D-69029,
Heidelberg, Postfach 103980, Germany\\
              \email{Heinrich.Voelk@mpi-hd.mpg.de}
         \and
             Institute for Terrestrial Magnetism, Ionosphere and Radiowave
Propagation, 142190, Troitsk, Moscow Region, Russia\\
             \email{Zirak@izmiran.rssi.ru}
             }

\abstract{

   Cosmic ray acceleration by shocks related with Slipping Interaction
Regions (SIRs) in the Galactic Wind is considered. SIRs are similar to Solar Wind
Corotating Interaction Regions. The spiral structure of our
Galaxy results in a strong nonuniformity of the Galactic Wind flow and in
SIR formation at distances of 50 to 100 kpc. SIRs are not corotating with
the gas and magnetic field because the angular velocity of the spiral
pattern differs from that of the Galactic rotation. It is shown that the
collective reacceleration of the cosmic ray particles with 
charge $Ze$ in the resulting
shock ensemble can explain the observable cosmic ray spectrum beyond the
"knee" up to energies of the order of $10^{17}Z$ eV.  
For the reaccelerated particles the Galactic Wind
termination shock acts as a reflecting boundary.

}

\maketitle
\section{Introduction}

The hypothesis that the origin of the cosmic rays (CRs) is predominantly the
result of diffusive shock acceleration at the blast waves of individual
supernova remnants (SNRs) has the corollary (e.g. Lagage \& Cesarsky \cite
{lagage}; V\"olk \cite{voelk87}; Axford \cite{axford}) that this mechanism
should only work up to particle rigidities (energy per charge) that are a
decade below the so-called knee of the spectrum at rigidities around several 
$10^{15}$ Volts, or somewhat more. The only consistent way to escape this
conclusion is to assume that the magnetic field strength at these shocks
substantially exceeds the typical interstellar values of several $\mu$G.
Such high field configurations might be unusually strong stellar fields in
the winds of very massive stars (Berezinsky \& Ptuskin \cite{berezinsky};
V\"olk \& Biermann \cite{voelkb}; Biermann \cite{biermann}), or the strong
Alfv\'{e}nic wave turbulence excited at strong shock waves, as speculated by
V\"olk (\cite{voelk84}) and recently calculated in a simplified nonlinear
model by Lucek \& Bell (\cite{lucek}) and Bell \& Lucek (\cite{belll}).
However, also this field amplification is limited and for a SN explosion
into an essentially uniform external medium (SNe type Ia and core collapse
SNe from stars of zero age main sequence masses below about 15 $M_{\odot}$
which have little main sequence mass loss) knee energies constitute a firm
upper bound (V\"olk et al. \cite{voelkbkr}). Our working hypothesis is
therefore that the knee corresponds to the cutoff for the CR sources in the
disk. Regarding core collapse SN explosions into the winds of massive
progenitor stars on the other hand, Bell \& Lucek (\cite{belll}) have even
pondered the possibility for an extension of the SNR origin of CRs to
energies far above the knee, up to $10^{18}$ eV for protons and to $10^{18}\
Z$ eV for heavy nuclei. The observed steepening of the all-particle energy
spectrum -- and presumably also of the energy spectra of individual nuclei
(Kampert et al. \cite{kampert}) -- above the knee at $3 \times 10^{15}$ Volt
they attributed somewhat vaguely to source statistics. Whether such a
picture of disk sources, with single power law spectra reaching beyond the
knee up to rigidities of about $10^{17}$ Volt, is consistent with the
rigidity dependence of the escape from the Galaxy, is an open question. In
any case no one of the very optimistic SNR scenarios aims at an explanation
of the all-particle spectrum beyond $3\times 10^{18}$ eV, the so-called
ankle, where the spectrum appears to harden again (for a recent summary, see
Sommers \cite{sommers}). And these ultra-high energy CRs are not our concern
here either. They may be extragalactic or due to a top-down mechanism of the
decay of hypothetical superheavy particles left over from the early Universe
(for a review see Bhattacharjee \& Sigl \cite{bhattacharjee}).

Assuming from now on that the CR sources in the disk cut off at the knee, we
propose here a rather different mechanism to accelerate CRs in the Galaxy to
energies between the knee and the ankle. This mechanism remains nevertheless
intimately connected with the mechanical energy input from stars into the
interstellar medium in the form of winds and SN explosions. For the process
to operate we start from the assumption that the Galaxy exhibits a
significant supersonic mass loss. It is mainly driven by the CRs and the hot
gas generated in the disk, and removes about $1 M_{\odot}$ yr$^{-1}$ from
the Galaxy with a terminal speed of several 100 km s$^{-1}$ that is of the
order of the escape velocity (Ipavich \cite{ipavich}; Breitschwerdt et al. 
\cite{breitschwerdt91, breitschwerdt93}; Zirakashvili et al. \cite{zirak};
Ptuskin et al. \cite{ptuskin}). Such a \textit{Galactic Wind} must develop
because these CRs are tightly coupled to the thermal gas and the magnetic
field of the interstellar medium and therefore tend to establish an
essentially infinite scale height above the disk due to their
ultrarelativistic mean energy. The coupling is due to scattering on magnetic
fluctuations which the escaping particles excite themselves in the
low-density halo of the Galaxy. Supersonic speeds are eventually reached at
distances of about 20 kpc corresponding to the size of the disk.

The new aspect which we emphasize here is based on the consideration that
even on a large scale this outflow must be far from regular in space and
time, and that its variations should be of the same order as the mean. We
shall argue that due to the rotation of the Galaxy the differences in flow
speed will lead to strong internal wind compressions, bounded by \textit{\
smooth} CR shocks in an expanding halo gas whose internal energy density is
dominated by that of the CRs from the disk. These CR shocks \textit{%
reaccelerate} the most energetic particles produced in the disk by about 2
orders of magnitude in rigidity. The re-acceleration essentially ensures the
continuity of the energy spectrum at the knee. A fraction of the
re-accelerated particles will return to the disk, filling a very thick
(several tens of kpc) region including the Galactic mid-plane rather \textit{%
uniformly and isotropically}. Since at these energies the propagation in the
average wind environment is diffusive (Ptuskin et al. \cite{ptuskin}), also
their spatial density in the disk is about equal to that of the knee
particles. We shall call the reaccelerated particles here "Wind-CRs", and
the primary CRs from compact regions in the disk we shall call "Disk-CRs",
for obvious reasons. These names indicate the origin of the particles which
is not only different in time but also in space.

Galactic reacceleration mechanisms have been frequently considered in the
past. This concerned reacceleration of CRs by shocks in the Galactic disk
(e.g. Blandford \& Ostriker \cite{blandford}, Berezhko \& Krymsky \cite
{berezhko}, Bykov \& Toptygin \cite{bykov}, Ip \& Axford \cite{ip}, Klepach
et al. \cite{klepach}), or acceleration of particles at the termination
shock of the Galactic Wind (Jokipii \& Morfill \cite{jokipii85},\cite
{jokipii87a} ).

Reacceleration in the Galactic disk can not give large enough energies of
accelerated particles because the size and life time of the shocks are quite
limited. A different case is the acceleration at the Galactic Wind
termination shock which can be steady state and is very extended (about
several hundred kpc). The investigation of acceleration on such a shock
results in rather large maximum energies of the accelerated particles of
about $10^{17} \ \mathrm{\ to}\ 10^{18}$ eV for CR protons (Jokipii \&
Morfill \cite{jokipii87a}) On the other hand, this form of reacceleration
faces the difficulty of the observation of these particles in the disk. The
matter is that the condition of efficient acceleration at the termination
shock coincides with the condition for strong modulation of particles inside
the Galactic Wind flow. For a diffusion coefficient which increases with
energy only the highest energy particles accelerated at the termination
shock can be observed near Earth.

\begin{figure}[tbp]
\centering
\includegraphics[width=7.0cm]{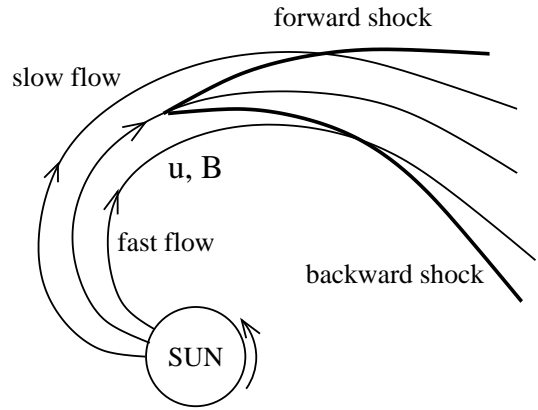}
\caption{Schematic picture of a Corotation Interaction Region in the Solar
Wind. The fast stream flows into the interaction region through the backward
shock, whereas the slow stream enters the interaction region through the
forward shock. Magnetic field and plasma velocity are parallel in the
reference frame rotating with the Sun.}
\label{Fig1}
\end{figure}

The alternative source of shocks in the Galactic Wind flow which we have
indicated above are interaction regions far above the rotating disk. They
are somewhat similar to the so-called Corotating Interaction Regions in the
Solar Wind (Fig. 1), although the differences in their characteristics are
quite important. Galactic interaction regions are due to the fact that much
of the high-mass star formation in the Galaxy and the associated active
regions such as superbubbles and OB- associations, including most of the
SNRs, is concentrated in the spiral arms. These rotate relative to the gas
in a quasi-stationary pattern with a related mechanical energy input from
stellar winds and SN explosions that implies an enhanced wind in terms of an
increased mass velocity (see also Breitschwerdt et al. \cite{breitschwerdt02}
). In the Parker spiral-type magnetic field
structure (Parker \cite{parker}) that the outflow from a magnetized Galactic
disk generates (Zirakashvili et al. \cite{zirak}), faster wind streams from
more active regions will begin to overtake slower streams after the wind has
formed in the upper halo. If the relative flow speed is supersonic, the
interaction will lead to a shock pair like in the Solar Wind. As long as the
flow speed of the wind is still low for a given rotation rate of the wind
source region, quasi-periodic compression regions will form that lead only
to a series of forward shocks after a steepening time, interleaved with
rarefaction waves. This distinct situation applies to the Galaxy, as we
shall see. Indeed, the flow time to the sonic point, from a reference level $%
\sim 1$~kpc above the disk midplane where the smooth wind flow hydrodynamics
is already a good approximation, is larger than or comparable with the
Galactic rotation time. In this sense the Galaxy is a fast rotator
whereas the Sun is a slow rotator: The wind flow time from
the base of the corona to the sonic point is short compared to the solar
rotation time\textbf{\textit{.}} A second difference is given by the fact
that the magnetic field lines are rooted in the gas of the disk, especially
in the massive molecular clouds, rather than in its spiral pattern, and are
therefore only temporarily compressed in the spiral arms. The angular
velocity of the interaction regions equals the spiral pattern angular
velocity and the angular velocity of frozen-in magnetic field lines equals
the Galactic angular velocity.


As a result, the interaction regions propagate as a wave phenomenon across
the magnetic flux tubes. From the point of view of the interaction regions
the magnetic field lines slip through them. That is why we shall rather
call these interaction regions Slipping Interaction Regions (SIRs). This
slip also means that particles that were accelerated at one of the shocks
will be able to leave the shock region eventually to escape along a field
line, rather than only across the field like in the Corotating Interaction
Regions of the Solar Wind. To this extent the acceleration properties of
the rotating Galactic halo are quite different from those of the
Corotating Interaction Regions of the Solar Wind discussed many years ago
by Fisk \& Lee (\cite{fisk}).

In this paper we shall consider reacceleration of Disk-CRs by SIR shocks in
the Galactic Wind flow, the large scale propagation of the high energy
Wind-CRs in the wind-disk regions, and the resulting anisotropies. The paper
is organized as follows: Brief descriptions of CR transport and of the
Galactic Wind flow are given in Sections 2 and 3, respectively. Numerical
calculations of SIR shock formation are presented in Section 4. CR
reacceleration by shocks related with the Galactic spiral structure is
introduced in Section 5 and calculated in Section 6. Section 7 contains the
conclusions.

\section{Cosmic ray propagation in the Galactic Wind flow}

\begin{figure}[tbp]
\centering
\includegraphics[width=7.0cm]{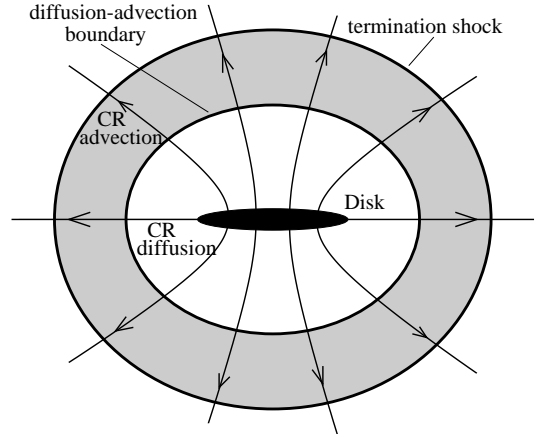}
\caption{Meridional cross-section of the Galactic Wind flow. The direction
of the gas velocity is shown by the arrows. The supersonic gas flow is
bounded by the termination shock. CR transport is mainly diffusive inside
the diffusion-advection boundary; outside this boundary it is determined by
convection in the gas flow (in the dashed region). The galactic disk is
indicated by the black ellipse. }
\label{Fig2}
\end{figure}

A schematic picture of the Galactic Wind flow geometry is given in Fig.2.
The azimuthally symmetric flow originates in the Galactic disk and extends
the frozen-in magnetic field. At small heights above the disk the gas
velocity is perpendicular to the galactic disk. The flow is approximately
radial at large distances from the Galaxy. The gas is assumed to be fully
ionized in the wind, supporting magnetohydrodynamic waves. These may also be
resonantly excited by the anisotropic streaming of the CR component. Their
amplitudes will ultimately be limited by damping through nonlinear
wave-particle interactions.

A complete and self-consistent model of CR propagation in such a flow has
been considered by Ptuskin et al. (\cite{ptuskin}). The CR diffusion
coefficient is determined by Alfv\'{e}n waves generated by the CR streaming
instability. CRs produced in the disk -- the Disk-CRs -- diffuse along
magnetic field lines. This diffusion flux generates Alfv\'{e}n waves and
they propagate in the Galactic Wind flow. The generation of Alfv\'{e}n waves
is balanced by nonlinear Landau damping. The resulting diffusion coefficient
$D_{\parallel }^{\mathrm{s}} $ along the magnetic field does not depend on
distance and is approximately given by

\begin{equation}
D_{\parallel }^{\mathrm{s}} \simeq 10^{27}(p/m_pc)^{\gamma _{\mathrm{d}} -3} 
\mathrm{cm^{2} s}^{-1}.  \label{diff}
\end{equation}

Here $p$ is momentum of the particle, $m_{\mathrm{p}}$ is the proton mass
and $\gamma _{\mathrm{d}} $\ is the power law index of the momentum
distribution of the CR sources in the disk; its numerical value is close to
4. The CR spectrum near in and near to the Galactic disk is formed as the
result of the interplay between diffusion and advection. At small distances
from Galactic midplane diffusion dominates advection. At large distances
diffusion is weaker because the gradient scale increases and because the
magnetic field becomes rather azimuthal. Diffusion and advection are
comparable at some distance $R_{\mathrm{da}} $ . This distance increases
with particle momentum as \ $p^{(\gamma _{\mathrm{d}} -3)/3}$\ . For the
diffusion coefficient given by Eq. (1) and a wind velocity $u=300$ km s$%
^{-1} $, $R_{\mathrm{da}} (1$TeV$)\approx 15$\ \ kpc, approximately equal to
the disk radius.

The supersonic Galactic Wind flow is bounded by a termination shock at some
distance $R_{\mathrm{s}} $ from the Galaxy. This distance depends on the
intergalactic pressure which is poorly known. Estimates give a value of
several hundred kpc. One has to expect that the termination shock creates
strong MHD turbulence downstream, towards intergalactic space. CR diffusion
is strongly reduced in this case and should be close to the Bohm limit. For
those CRs whose propagation characteristics in the Galactic Wind (inside the
termination shock) are diffusive, the termination shock can therefore be
considered as a reflecting boundary. This holds up to a maximum particle
rigidity that is determined by the condition $D_{\mathrm{B}} \sim uR_{%
\mathrm{s}} $\, where $D_{\mathrm{B}}=vr_{\mathrm{g}}/3$ \ is the Bohm
diffusion coefficient of particles with gyroradius $r_{\mathrm{g}}$ and
velocity $v$. Its numerical value is determined by the local magnetic field
strength. At large distances in the then spherically symmetric Galactic Wind
flow the magnetic field strength is given by

\begin{equation}
B\approx B_{\mathrm{g}}\frac{R_{\mathrm{g}}^{2}\Omega (\theta )\sin \theta }{%
Ru},
\end{equation}
where $\theta $\ \ denotes Galactic colatitude, $B_{\mathrm{g}}$~ is the
poloidal field strength in the disk, $R_{\mathrm{g}}$\ is the Galactic
radius and \ $\Omega $\ is the angular velocity of Galactic rotation. Then,
using Eq. (2), the above estimate \textit{$D_{\mathrm{B}}\sim 
uR_{\mathrm{s}}$\ }
gives the following maximum energy

\[
E_{\max } = Z\sin \theta \left( \frac{\Omega (\theta )}{10^{-15} \mathrm{s}
^{-1}} \right) \left( \frac{B_{\mathrm{g}} }{2\mu \mathrm{G}}\right) \cdot 
\]
\begin{equation}
\left( \frac{R_{\mathrm{g}} }{15\mathrm{kpc}}\right) ^{2}1.2\cdot 10^{17} 
\mathrm{eV},  \label{Emax}
\end{equation}
where $Z$\ is the nuclear charge number of the particle. We should note that
this energy is not small for small $\theta $ because the angular velocity of
the galactic rotation $\Omega $\ increases toward the galactic center.
Higher energy particles cross the termination shock diffusively.

\subsection{Single power law sources up to $E_{\max }$ in the disk?}

Let us assume now that the sources in the Galactic disk produce CRs with a
single power law spectrum up to very high rigidities. If the rigidity of the
particles is small enough so that the distance to the diffusion-advection
boundary $R_{\mathrm{da}} (p) < R_{\mathrm{s}} $ then in the Galactic disk
the spectrum of such Disk-CRs will be modified relative to the source
spectrum proportional to $p^{-\gamma _{\mathrm{d}} }R_{\mathrm{da}} ^{-2}(p)$%
\ . This happens as long as $E<E_{\ast }$\ , where \ $E_{\ast }$\ \ is given
by $R_{\mathrm{da}} (E_{\ast })=R_{\mathrm{s}} $\ and corresponds to \ $%
E_{\ast }\sim 10^{16}Z$ eV for $R_{\mathrm{s}} =300$ kpc\ . Particles with
larger rigidities freely diffuse in the space interior to the termination
shock and can only escape convectively through this shock in an energy
independent manner because of the small diffusion coefficient beyond. Hence
the observed spectrum must be the same as the source spectrum. As a result
we should observe a hardening of the spectrum in the energy range $E_{\ast
}<E<E_{\max }$\ (see Appendix A for an approximate analytical calculation of
this effect).

However, this spectral hardening is not observed. We therefore conclude that
Galactic sources must have a cut-off energy smaller than $E_{\ast }$\ which
in turn is essentially equal to the knee energy \ $E_{\mathrm{knee}}\simeq
3\cdot 10^{15}$\ eV. As a consequence the notion of a single power law
source spectrum, produced in the Galactic disk, must be ruled out. Its basis
was the somewhat arbitrary assumption that Galactic escape might become even
more effective beyond the knee (e.g. Hillas \cite{hillas}). In the light of
a consistent physical model for the overall CR transport in the Galaxy this
hypothesis is untenable. We rather believe that reacceleration in the
Galactic Wind inside the termination shock should produce particles with
energies $\ E_{\mathrm{knee}}<E<E_{\max }$\ with the same source spectrum as
the spectrum observed in the disk. In a bottom-up scenario of acceleration
from lower energies, higher energy particles should mainly come from beyond
the termination shock.

\subsection{Quantitative definition of Disk-CRs vs. Wind-CRs}

We can now make our qualitative definitions of Disk-CRs and Wind-CRs from
the Introduction more precise: equating $E_{\mathrm{\ knee}}=E_{\ast }$\,
the Disk-CRs are those energetic particles with $E<E_{\mathrm{knee}}$, and
their sources are in the disk; Wind-CRs are the particles with $E_{\mathrm{\
knee}}<E<E_{\max }$\ whose source is of a quasi-diffuse nature in the
Galactic Wind.

\section{Spiral structure of our Galaxy, wind shocks}

The spiral pattern of late-type galaxies was explained theoretically by Lin
and Shu (\cite{lin}) as a density wave in the distribution of stars in the
disk of galaxies. It is believed now that our Galaxy possesses a two or four
spiral arm structure, or even a mixed two-four arm spiral pattern
(L\'{e}pine et al. \cite{lepine}). This pattern rotates rigidly with angular
velocity $\Omega _{\mathrm{p}}$. Therefore the dependence of the wave
amplitude on time $t$ and azimuthal angle $\varphi $ is given by $\exp ( 
\mathrm{i}m(\varphi -\Omega _{\mathrm{p}}t))$, where $m$ is the number of
spiral arms. The estimates for the angular velocity of the spiral pattern
are very uncertain and controversial. The old value is $\Omega _{\mathrm{p}
}=13.5$ km s$^{-1}$ kpc$^{-1}$ (Lin et al. \cite{LYS}). L\'{e}pine and
Amaral (\cite{la}) found $\Omega _{\mathrm{p}}=20$ km s$^{-1}$ kpc$^{-1}$,
L\'{e}pin et al. (\cite{lepine}) argue in favor of $\Omega _{\mathrm{p}}=26$
km s$^{-1}$ kpc$^{-1}$. On the other hand Fernandez et al. (\cite{fernandez}
) give the value $\Omega _{\mathrm{p}}=30$ km s$^{-1}$ kpc$^{-1}$. In any
case, the Sun is not located far from the corotation radius at which the
Galactic rotation velocity equals the pattern velocity. The rotation
velocity of the Galaxy at the Sun's position is approximately $\Omega =26$
km s$^{-1}$ kpc$^{-1}$.

Since potential CR sources in the disk are primarily concentrated in the
spiral arms, we can assume that the CR pressure in and above the spiral arms
is larger than between the arms. Hence, the CR-driven Galactic Wind flow
should be modulated by the spiral structure. The situation differs from the
Solar Wind case because of the relatively fast rotation of the Galaxy.
Indeed, the Galactic Wind flow time to the (magneto)sonic point is about 100
million years and is comparable with the period of Galactic rotation. This
means that the modulation by the spiral pattern should produce magnetosonic
waves, propagating in the Galactic Wind flow. At large distance from the
Galaxy these spiral compression waves propagate in the radial direction and
approximately perpendicular to the wind magnetic field which is by then
practically azimuthal. The radial and azimuthal wavenumbers for these waves
are

\begin{equation}
k_{r}=\frac{\Omega _{\mathrm{p}}m}{u_{\mathrm{s}} },k_{\phi }=\frac{m}{r},
\label{k}
\end{equation}
where $u_{\mathrm{s}} $ the radial velocity of the wave. This value is
approximately the sum of the wind velocity $u$\ and the phase velocity of
the fast magnetosonic waves $c_{\mathrm{f}} $\ traveling in the radial
direction. At large distances we simply have

\[
c_{\mathrm{f}} =\sqrt{V_{\mathrm{a}} ^{2}+c_{\mathrm{s}} ^{2}} 
\]
where $V_{\mathrm{a}} $ and $c_{\mathrm{s}} $ are the Alfv\'{e}n and the
sound velocity, respectively.

The nonlinear steepening of magnetosonic waves can produce a train of
forward shocks at large Galactocentric distances. The characteristic
distance can be found from the following estimate. The inverse steepening
time is $k_{r}\delta u$, where $\delta u$ is the velocity perturbation in
the wave. During this time the wave propagates the distance $\delta r\sim
(u+c_{\mathrm{f}} )/(k_{r}\delta u)$. Using the expression for the radial
wave number we obtain

\[
\delta r\sim \frac{(u+c_{\mathrm{f}} )^{2}}{m\Omega _{\mathrm{p}}\delta u} 
\]
For $u+c_{\mathrm{f}} \sim $400 km s$^{-1}$ , $\delta u\sim $ 50 km s$^{-1}$
, and $m=2$, $\delta r$ is about 60 kpc. The velocity perturbation $\delta u$
is half of the velocity jump at the shock formed. We therefore conclude that
spiral shock formation is possible at distances of 60 to 100 kpc. Numerical
results (see below) confirm this estimate. These spiral shocks should not be
confused with the spiral density wave in the Galactic disk. The radial
dependence of the spiral density wave in the Galactic disk will be
transformed into a latitude dependence of the spiral shocks in the Galactic
Wind flow (see Fig.2).

We should underline that the shocks in the Galactic Wind have two distinct
features in comparison with the Solar Wind.

First of all the shock structure is stationary in the frame of reference
corotating with the spiral pattern. Material Galactic rotation still exists
in this frame except at special Galactocentric radii. Therefore these shocks
are not in corotation with the matter in the Galactic disk but rather slip
through it and this is why we shall call them SIR shocks. Such a particular
feature can be very important for diffusive shock acceleration of particles.
This can be appreciated by comparing the Galaxy with the Sun. It is simple
to see that the particles accelerated in the shocks of Corotating
Interaction Regions of the Solar Wind are drawn into the Corotating
Interaction Region between the forward and backward shocks (see Fig.1). In
addition, energetic particles diffuse predominantly along magnetic lines
which are immobile in the frame corotating with the Sun. Hence, particles
that are moving backward to the Sun should pass through the shock again,
which is difficult because of the small diffusion coefficient generated near
the shock and in the downstream region of the gas flow. Therefore it is only
possible to observe these particles in the space between the Sun and the
Interaction Region (e.g. in the neighborhood of the Earth) if they diffuse
across the field which is a very slow process. Galactic SIR shocks, on the 
other hand, are not in
corotation with the wind magnetic field lines because these field lines are
anchored in the Interstellar gas of the Galactic disk. Hence the inner parts
of field lines become ''free'' towards the disk (after the shock has left
them behind and the shock-created turbulence has subsided). This allows
SIR-accelerated particles to reach the disk along magnetic field lines
without any necessity to invoke perpendicular diffusion.

The second feature is the absence of backward SIR shocks. Outward moving
large-amplitude periodic waves steepen into a train of shocks (saw-tooth
wave), where forward shocks are followed by rarefaction waves, like in a gas
at rest.

\section{Numerical calculations of SIR shocks}

In this Section we present numerical calculations of SIR shock formation in
the Galactic Wind flow. We shall use a simplified spherical geometry for the
Galactic Wind flow and investigate the formation of SIR shocks propagating
in the Galactic equatorial plane. For simplicity we describe thermal gas and
CRs in a two-fluid approximation with two-dimensional magnetohydrodynamic
equations including CRs (see e.g. Zirakashvili et al. (1996)):

\begin{equation}
\frac{\partial \rho }{\partial t}=-\frac{1}{r^{2}}\frac{\partial }{\partial
r }\left( r^{2}u\rho \right) -\frac{1}{r}\frac{\partial }{\partial \varphi }
\left( u_{\varphi }\rho \right)  \label{ro}
\end{equation}

\[
\frac{\partial u}{\partial t}=-u\frac{\partial u}{\partial r}-\frac{
u_{\varphi }}{r}\frac{\partial u}{\partial \varphi }+\frac{u_{\varphi }^{2}}{
r}-\frac{\partial \Phi }{\partial r}- 
\]
\begin{equation}
-\frac{1}{\rho }\frac{\partial }{\partial r}\left( P_{\mathrm{c}} +P_{%
\mathrm{g}} +\frac{ B_{\varphi }^{2}}{8\pi }\right) +\frac{1}{4\pi \rho }%
\left( \frac{B_{\varphi }}{r}\frac{\partial B}{\partial \varphi }-\frac{%
B_{\varphi }^{2}}{r}\right)  \label{u}
\end{equation}

\[
\frac{\partial u_{\varphi }}{\partial t}=-u\frac{\partial u_{\varphi }}{
\partial r}-\frac{u_{\varphi }}{r}\frac{\partial u_{\varphi }}{\partial
\varphi }-\frac{u_{\varphi }u}{r}- 
\]
\begin{equation}
-\frac{1}{\rho r}\frac{\partial }{\partial \varphi }\left( P_{\mathrm{c}}
+P_{\mathrm{g}} +\frac{ B^{2}}{8\pi }\right) +\frac{1}{4\pi \rho }\left( B
\frac{\partial B_{\varphi } }{\partial r}+\frac{B_{\varphi }B}{r}\right)
\label{uf}
\end{equation}

\begin{equation}
\frac{\partial B_{\varphi }}{\partial t}=-u\frac{\partial B_{\varphi }}{
\partial r}-\frac{u_{\varphi }}{r}\frac{\partial B_{\varphi }}{\partial
\varphi }+Br\frac{\partial }{\partial r}\frac{u_{\varphi }}{r}-\frac{
B_{\varphi }}{r}\frac{\partial }{\partial r}\left( ru\right)  \label{bf}
\end{equation}

\begin{equation}
\frac{1}{r}\frac{\partial B_{\varphi }}{\partial \varphi }+\frac{1}{r^{2}} 
\frac{\partial }{\partial r}\left( r^{2}B\right) =0  \label{div}
\end{equation}

\[
\frac{\partial P_{\mathrm{g}} }{\partial t}=-u\frac{\partial P_{\mathrm{g}} 
}{\partial r}-\frac{ u_{\varphi }}{r}\frac{\partial P_{\mathrm{g}} }{%
\partial \varphi }-\gamma _{\mathrm{g}} P_{\mathrm{g}} \left( \frac{1}{r^{2}}
\frac{\partial }{\partial r}\left( r^{2}u\right) +\frac{1}{r}\frac{\partial
u_{\varphi }}{\partial \varphi } \right) 
\]
\begin{equation}
-\left( \gamma _{\mathrm{g}} -1\right) \left( v_{\mathrm{a}} \frac{\partial
P_{\mathrm{c}} }{\partial r} +v_{\mathrm{a}\varphi }\frac{\partial P_{%
\mathrm{c}} }{\partial \varphi }\right)  \label{pg}
\end{equation}

\[
\frac{\partial P_{\mathrm{c}} }{\partial t}=-(u+v_{\mathrm{a}} )\frac{
\partial P_{\mathrm{c}} }{\partial r} -\frac{u_{\varphi }+v_{\mathrm{a}
\varphi }}{r}\frac{\partial P_{\mathrm{c}} }{\partial \varphi } - 
\]
\begin{equation}
-\gamma _{\mathrm{c}} P_{\mathrm{c}} \left( \frac{1}{r^{2}}\frac{\partial }{
\partial r}\left( r^{2}(u+v_{\mathrm{a}} )\right) +\frac{1}{r}\frac{\partial 
}{\partial \varphi } (u_{\varphi }+v_{\mathrm{a}\varphi })\right)  \label{pc}
\end{equation}
Here $\rho $ is the gas density, $u$\ and $B$ denote the radial components
of the gas velocity and the magnetic field respectively, $u_{\varphi }$ and $%
B_{\varphi }$ are the azimuthal components of the gas velocity and the
magnetic field, $\Phi $ is the gravitational potential, $P_{\mathrm{g}} $
and $P_{\mathrm{c}} $ are the gas and CR pressures, respectively, $\gamma _{%
\mathrm{g}} $ and $\gamma _{\mathrm{c}} $ with $\gamma _{\mathrm{c}} <
\gamma _{\mathrm{g}} $ are the adiabatic indices of the thermal gas and the
CR gas, and $v_{\mathrm{a}} =B/\sqrt{4\pi \rho }$ and $v_{\mathrm{a}\varphi
}=B_{\varphi }/\sqrt{4\pi \rho }$ are the radial and azimuthal components of
the Alfv\'{e}n velocity, respectively.

Eq. (\ref{ro}) describes mass conservation. The overall momentum balance
Eqs. (\ref{u}) and ( \ref{uf}) determine the mass motion in the radial and
azimuthal directions. Eqs. (\ref{bf}) and (\ref{div}) describe the evolution
of the frozen-in magnetic field. The time evolution of the CR pressure is
given by Eq. (\ref{pc}), neglecting diffusion in the frame of the waves.
This is an excellent approximation for the description of these large-scale
structures\textbf{\textit{\ %
%
}}since the CR pressure is mainly determined by low-energy CR particles
produced in the Galactic Disk. The transport of these particles is
completely dominated by advection in the SIR waves\textbf{\textit{.}}
Following Zirakashvili et al. ( \cite{zirak}) we assume that CR streaming
generates Alfv\'{e}n waves propagating along the magnetic field out of the
Galaxy. These waves are damped and heat the gas. This dissipative effect is
described by the last term in the equation for the gas pressure, Eq. (\ref
{pg}).

The gravitational potential $\Phi$ was chosen in the following simplified,
spherically symmetric form:

\begin{equation}
\Phi =\Phi _{0}-\frac{GM_{\mathrm{B,D}}}{r}+\frac{GM_{\mathrm{H}} }{R_{%
\mathrm{H}} }\ln \left( 1+\frac{r}{ R_{\mathrm{H}} }\right)
\end{equation}
for $r<100$ kpc. Here $G$=6.668$\cdot 10^{-8}$ cm$^{3}$g$^{-1}$s$^{-2}$
denotes the gravitational constant. The second term on the r.h.s. describes
the gravitational potential of the Galactic disk and bulge, \ and the third
term is the potential of the Dark Matter Halo with radius 100 kpc. We used
the values $M_{\mathrm{B,D}}=9.97\cdot 10^{10}M_{\odot }$, $M_{\mathrm{H}}
=1.07\cdot 10^{11}M_{\odot }$, $R_{\mathrm{H}}=12$ kpc from Allen \&
Santill\'{a}n (\cite{allen}). The total mass of the Dark Matter Halo amounts
to $8.0\cdot 10^{11}M_{\odot }$.

The system of equations (\ref{ro}) to (\ref{pc}) was solved numerically
using an explicit finite-difference scheme. Artificial viscosity was
included to avoid numerical instabilities. At the base level at $r=15$ kpc
we fixed the gas number density $n_{0}=0.001$ cm$^{-3}$, the radial magnetic
field component $B_{0}=1.0\cdot 10^{-6}$ G, the gas temperature $T=7.25\cdot
10^{5}$ K, and the azimuthal velocity $u_{\varphi 0}=220$ km s$^{-1}$ .
Adiabatic indices $\gamma _{\mathrm{g}} =5/3$ and $\gamma _{\mathrm{c}} =4/3$
were used. The CR energy flux $F_{\mathrm{c}} =\gamma _{\mathrm{c}} (u+v_{%
\mathrm{a}} )P_{\mathrm{c}} /(\gamma _{\mathrm{c}} -1)$, modulated by the
spiral structure, was taken in the form

\begin{equation}
F_{\mathrm{c}} =F_{\mathrm{c0}} \left( 1+0.5\sin (m(\varphi -\Omega _{%
\mathrm{p}}t))\right)
\end{equation}
which assumes that the power of the Disk-CR sources in the spiral arms is a
factor of three larger than between the arms. For the numerical calculations
we used $m=2$, $\Omega _{\mathrm{p}}=30$ km s$^{-1}$ kpc$^{-1}$ and $F_{%
\mathrm{c0}} =2.0\cdot 10^{-5}$ erg cm$^{-2}$ s$^{-1}$.

Numerical results are shown in Fig.3. One can see that SIR shocks with a
velocity jump of the order 100 km s$^{-1}$ are formed at distances exceeding
50 kpc. These SIR shocks form a saw-tooth wave velocity profile. The
compression ratio $\sigma _{\mathrm{s}} $ of these shocks is about 2.0. In
the sequel we shall call the entire saw-tooth wave the ''SIR shock system''.

\begin{figure*}[tbp]
\centering
\includegraphics[width=17.0cm]{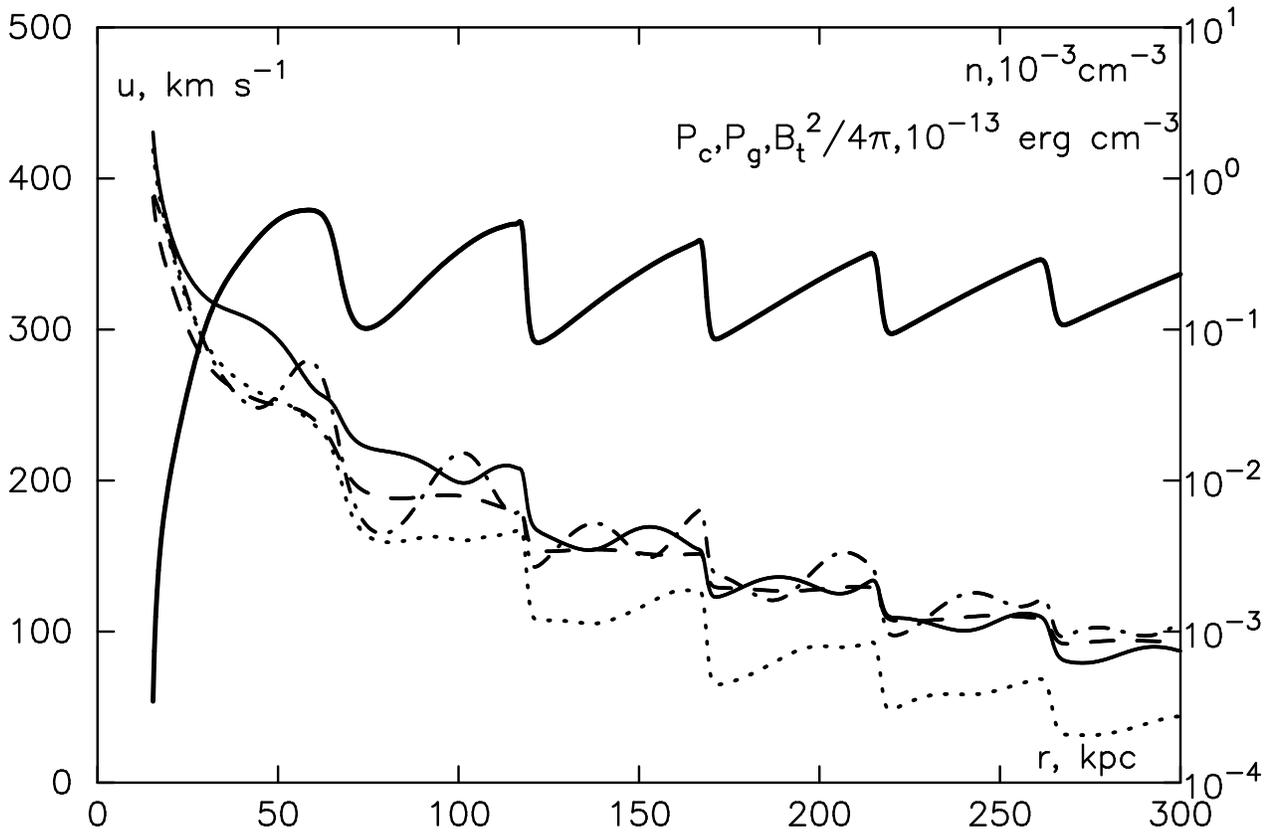}
\caption{Radial dependencies, taken at one azimuth angle. The values of the
radial gas velocity $u$ (thick solid line, in units of km s$^{-1}$ ) are
given on the left abscissa. The right abscissa shows the values of the
cosmic ray and gas pressures $P_{\mathrm{c}}$ (thing solid line) and $P_{%
\mathrm{g}}$ (dotted line), respectively (in units of $10^{-13}$ erg cm$%
^{-3} $), the gas number density $n$ (dashed line, in units of $10^{-3} 
\mathrm{cm}^{-3}$), and the total magnetic field tension $B^{2}_t/4\protect%
\pi $ (dash-dotted line, in units of $10^{-13}$ erg cm$^{-3}$). Forward SIR
shocks form a saw-tooth velocity profile at large distances in the Galactic
Wind flow.}
\label{Fig3}
\end{figure*}

We emphasize that our numerical results show the presence of forward shocks
only. Backward shocks were also observed at initial times in the
computation. Nevertheless the backward shocks were blown out at large times
when all quantities depend on $r$ and $\varphi -\Omega _{\mathrm{p}}t$ only.
A similar result was obtained by Cranmer \& Owocki (\cite{cranmer}) for the
formation of Corotating Interaction Regions in O-star winds.

\begin{figure*}[tbp]
\centering
\includegraphics[width=17.0cm]{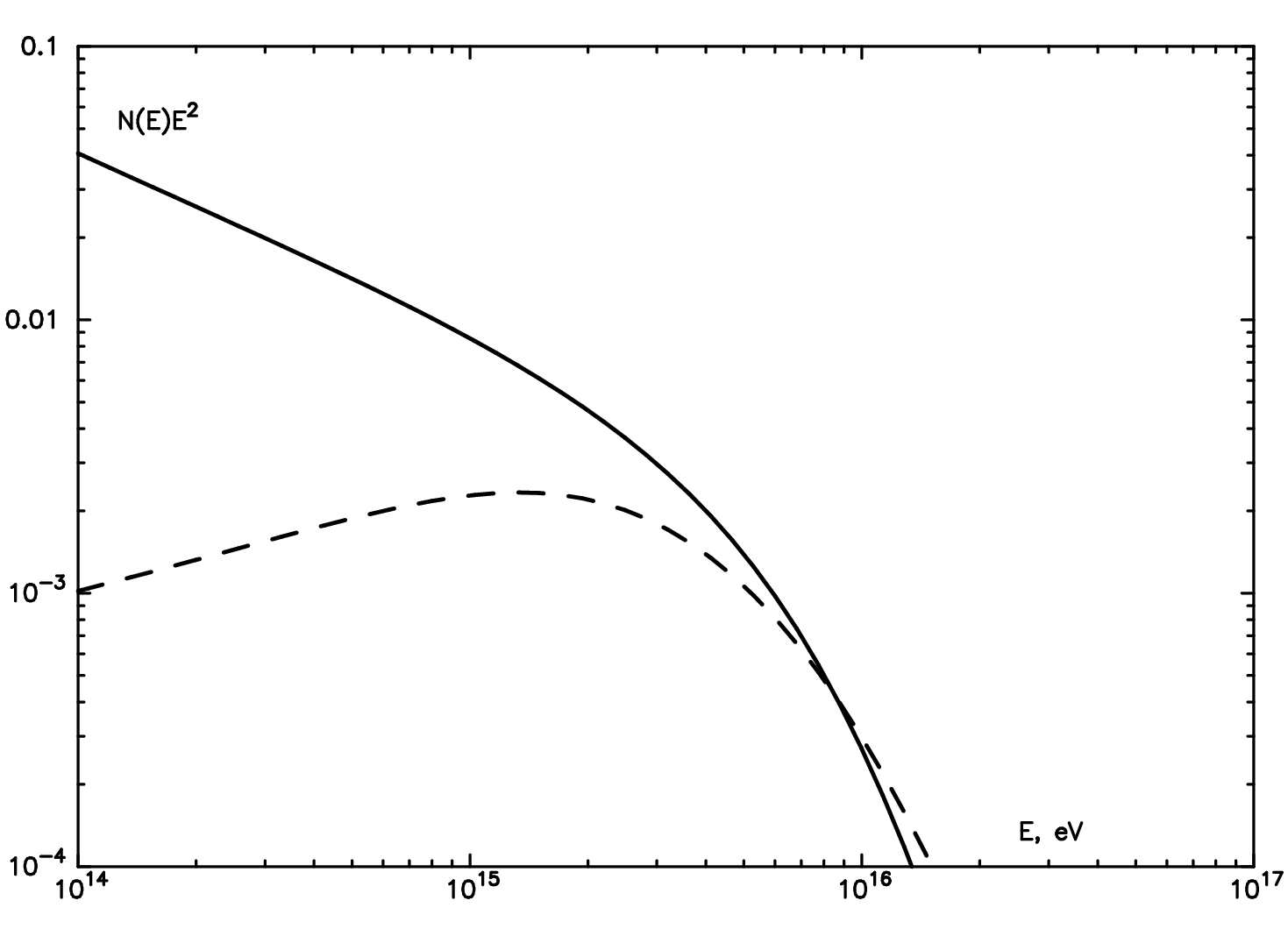}
\caption{Spectral energy distributions (in arbitrary units) of the CR
protons in the Galaxy (solid curve) and at the termination shock (dashed
curve) for the case without reacceleration.}
\label{Fig4}
\end{figure*}

\begin{figure*}[tbp]
\centering
\includegraphics[width=17.0cm]{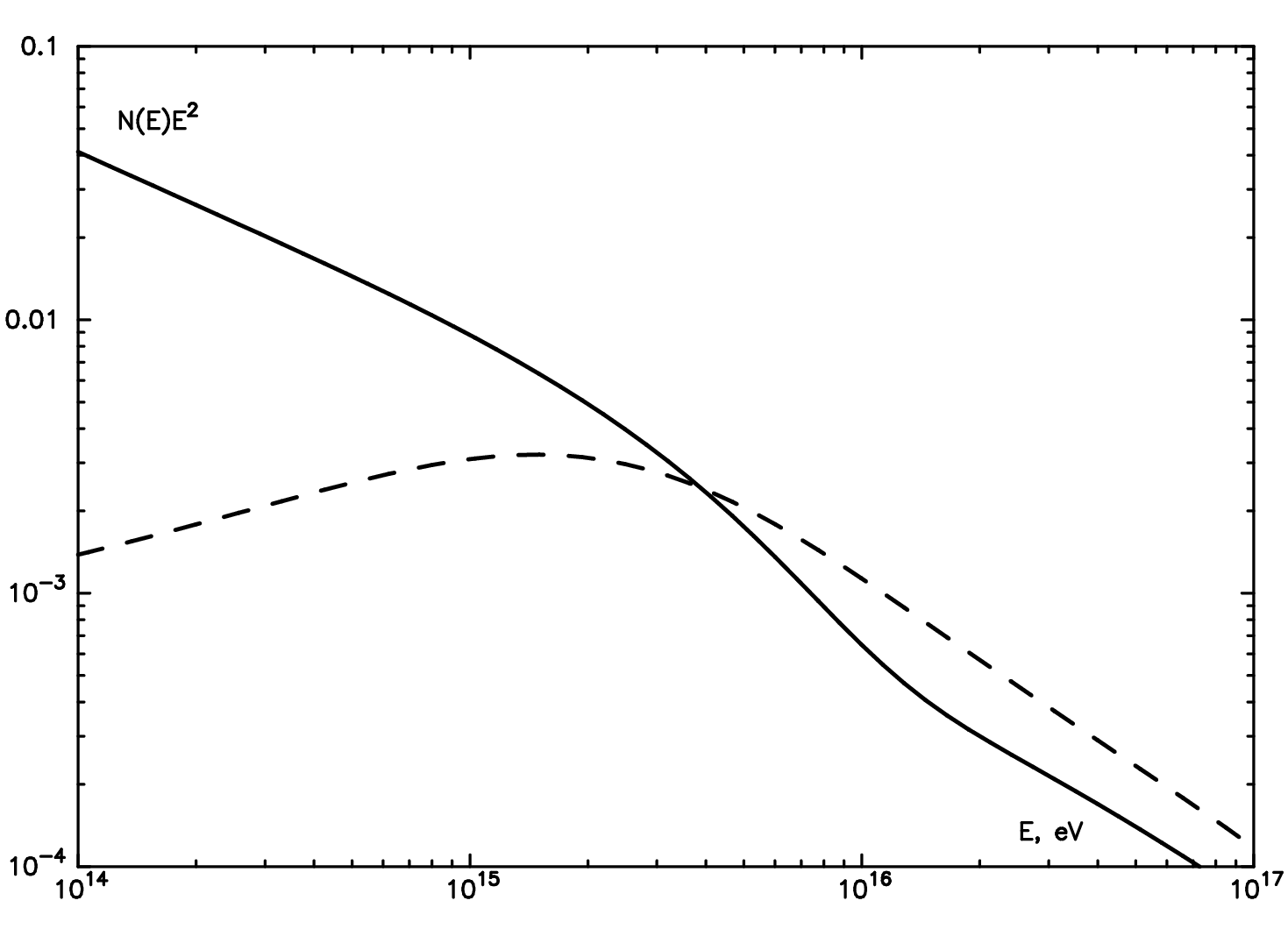}
\caption{Spectral energy distributions (in arbitrary units) of the CR
protons in the Galaxy (solid curve) and at the termination shock (dashed
curve) for the case with reacceleration.}
\label{Fig5}
\end{figure*}

\section{Reacceleration of particles by the SIR shock system: physics
conditions}

We shall consider now the possibilities for standard diffusive shock wave
acceleration in the spiral SIR shock system. A similar consideration was
made \ by Spruit (\cite{spruit}) for spiral accretion shocks. Due to the
fact that the wind is primarily driven by the CRs which have a softer
equation of state and dominate the thermal part of the pressure
(Zirakashvili et al. \cite{zirak}), Galactic Wind SIR shocks are basically
smoothed by the pressure of the ''low'' energy Disk-CRs (compare Drury \&
V\"olk \cite{drury}); in addition, they are almost perpendicular shocks. In
this case there is \textit{no injection of particles from the thermal pool}
and SIR shocks will only \textit{reaccelerate} Disk-CRs with energies $E$
close to $E_{\mathrm{knee}}$ to energies beyond the knee.

It is interesting to consider the implications of this situation. First of
all, the SIR shock system will not modify the spectral form of the Disk-CRs
below the knee. Since the average adiabatic compression in the saw-tooth
wave is zero, there will also be no net adiabatic number or energy density
increase of the Disk-CRs; the spectrum is still determined by the interplay
of the sources in the disk and the propagation in the wind. Secondly, the
energy requirements for the production of the Wind-CRs by the SIR shock
system are minimal because no other particles are energized in the process.
The total escape energy loss of the Wind-CRs is given by their convective
energy flow through a spherical surface at the radial position of the
termination shock (see below).

The magnetic field in the Galactic Wind is rather strong. As a result, the
Mach number of the wind flow is not very large. We expect that SIR shocks
are not very strong either and that the spectrum of the particles
accelerated by a single shock of the SIR shock system will be fairly steep.
On the other hand, particles with energies of the Wind-CRs are diffusively
locked inside the termination shock and can be continuously reaccelerated by
the system of SIR shocks. It is well known that in such a situation the
spectrum of accelerated particles can be harder than that due to a single
shock (e.g. Blandford \& Ostriker \cite{blandford}, Spruit \cite{spruit},
Achterberg \cite{achterberg}, Melrose \& Pope \cite{melrose}, Klepach et al. 
\cite{klepach}). The only caveat here is that all those many-shock
acceleration theories were based on the test particle approximation in a
system of shocks with a strictly discontinuous velocity profile. We shall
also follow this approach here, even though the finite width of the actual
SIR shocks should not only produce a power law distribution of Wind-CRs but
at the same time prevent any increase of the Disk-CRs.

The Parker spiral angle $\alpha $ between the magnetic field and the radial
direction is given by the expression

\begin{equation}
\cos \alpha =\frac{u}{\sqrt{\Omega ^{2}r^{2}\sin ^{2}\theta +u^{2}}}.
\label{alpha}
\end{equation}

We shall assume that the Galactic Wind flow is bounded by a strong
termination shock at distance $R_{\mathrm{s}} $, where the wind velocity
drops by a factor $\sigma $, the shock compression ratio. Beyond the
termination shock the gas flow is subsonic and is dynamically dominated by
the sum of gas, magnetic, and CR pressures. One may assume that the Galactic
Wind flow is highly turbulent in this downstream region. We shall assume
that for all particles of Galactic origin, Disk-CRs and Wind-CRs, diffusion
is negligible in comparison with advection beyond the termination shock.

The scattering of the energetic particles is weaker interior to the
termination shock. Nevertheless, as already mentioned in the Introduction,
particles with $E<E_{\max }$ can be accelerated at the termination shock if
the diffusion coefficient is small enough there. They can not be observed
inside the Galaxy for the same reason. Very high energy particles, with $%
E>E_{\max }$ and a correspondingly high diffusion coefficient, coming from
the termination shock or beyond, can be observed in the disk, but can not be
accelerated at the termination shock.

We proceed now to the calculation of the reacceleration of Disk-CRs by the
SIR shock system with spatial period $L=2\pi /k_{r}$ and velocity jump $%
\Delta u$. The condition for efficient diffusive shock wave acceleration is $%
D\lesssim (u_{\mathrm{s}}-u)L/6$, where $D$ is the diffusion coefficient in
the direction of the shock normal. It means that the diffusion time\textbf{\ 
} $(Dk_{r}^{2})^{-1}$must be large in comparison with the advection time%
\textbf{\ }$(k_{r}(u_{\mathrm{s}}-u))^{-1}$\textbf{. }In the Bohm limit this
condition gives a maximum energy of accelerated particles that is smaller
than the maximum energy given by equation (\ref{Emax}). However, the SIR\
shocks are almost perpendicular and therefore the diffusion coefficient in
the direction of the shock normal can be smaller than the Bohm value if the
shock-associated turbulence is moderate. This implies a larger maximum
energy than that corresponding to the Bohm limit. A necessary condition for
this situation to apply is $D_{\parallel }/D_{\mathrm{B}}<v/u_{\mathrm{s}}$
(Jokipii \cite{jokipii87b}). For our parameters this condition is fulfilled
within a large dynamic range. However expression (\ref{Emax}) still gives
the maximum energy for collective reacceleration , because at larger
energies the particles leave across the terminal shock rather than remaining
confined in the reacceleration region. Perpendicular collisionless shocks
with Mach numbers smaller than 2 should have a laminar structure (cf.
Sagdeev \cite{sagdeev}, Forslund \& Freiberg \cite{forslund}). Therefore a
moderate level of MHD turbulence downstream of the individual SIR shocks and
small CR diffusion coefficients in the direction of the shock normals are
possible. Drift motions in the inhomogeneous regular magnetic field in the
latitudinal direction can also diminish the maximum energy of accelerated
particles since the SIR shock system depends on latitude. However, we expect
the regular magnetic field to be small because the regular poloidal magnetic
field component is weak in our Galaxy (cf. Han \& Qiao \cite{han}). Galactic
Wind streams originating from different parts of the Galactic disk should
drag out random magnetic field and magnetic loops with sizes of the order of
1 kpc. These magnetic disturbances become strongly elongated at large
distances from the Galaxy due to the acceleration and spherical expansion of
the Galactic Wind flow (cf. Zirakashvili et al. \cite{zirak2}). Hence, the
distant wind is filled by almost azimuthal, sign-dependent \ magnetic
fields. But Parker's formula (\ref{alpha}) and similar expressions for the
magnetic field strength remain valid for this case as well. CR diffusion in
such a field should be very anisotropic. Random drift motions and wandering
of magnetic field lines would produce an anomalous transport across magnetic
lines.

In the following we shall concentrate on the problems of continuity of the
spectrum of the reaccelerated particles and of the possibility to observe
them in the Galaxy. For the sake of simplicity we shall neglect anomalous
perpendicular transport and treat CR diffusion as one-dimensional \textbf{\ }%
similar to the paper of Fisk and Lee (1980).\newline
Let us calculate\ the angle between the magnetic field and the shock normal $%
\alpha _{\mathrm{s}}$. Using (\ref{k}) and \ (\ref{alpha}) we obtain for $%
r>>u/\Omega $ \ 
\begin{equation}
\cos \alpha _{\mathrm{s}}=\left( \frac{\Omega u_{\mathrm{s}}}{\Omega _{%
\mathrm{p}}u}-1\right) \cos \alpha   \label{cos}
\end{equation}

One can easily see that this can be zero for a Galactic Wind flow line
originating at a particular galactocentric radius, which means that the
corresponding SIR shocks are purely perpendicular shocks. The acceleration
by these shocks will be efficient if \ $D_{\parallel }\cos ^{2}\alpha _{%
\mathrm{s}} \lesssim (u_{\mathrm{s}} -u)L/6$. On the other hand, particles
accelerated near the termination shock can be observed in the Galaxy if $%
D_{\parallel }\cos ^{2}\alpha \gtrsim uR_{\mathrm{s}} $. These two
inequalities can coexist only in the case

\begin{equation}
\left( \frac{\Omega u_{\mathrm{s}} }{\Omega _{\mathrm{p}}u}-1\right) ^{2}< 
\frac{(u_{\mathrm{s}} -u)L}{ 6uR_{\mathrm{s}} }=\frac{\pi (u_{\mathrm{s}}
-u)u_{\mathrm{s}} }{3mu\Omega _{\mathrm{p}}R_{\mathrm{s}} }.  \label{cond}
\end{equation}

This condition determines the part of the Galactic Wind flow filled by
effectively accelerating SIR shocks which produce CRs that are observable in
the Galactic disk. They are the Wind-CRs. 
\begin{figure*}[tbp]
\centering
\includegraphics[width=17.0cm]{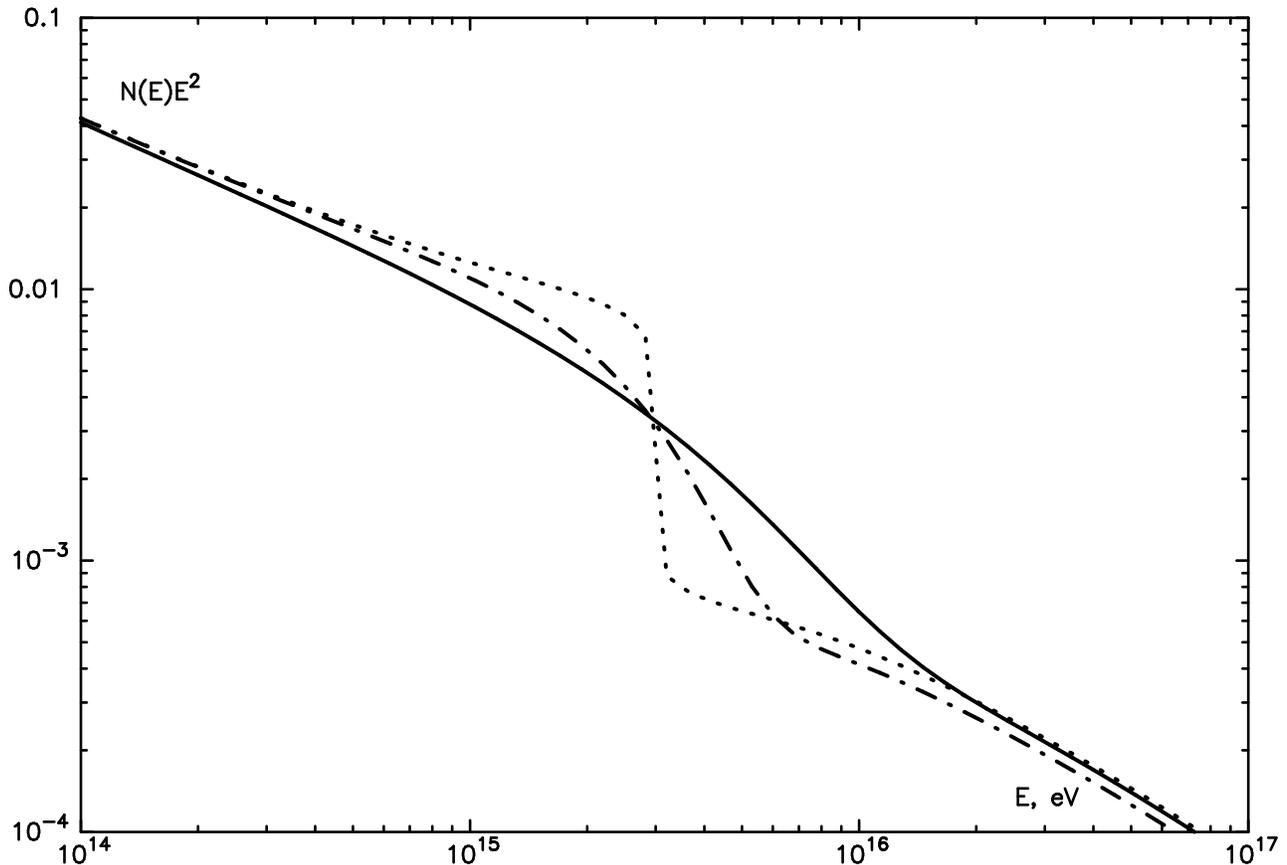}
\caption{Spectral energy distributions (in arbitrary units) of the CR
protons in the Galaxy for different cut-off forms: exponential (solid
curve), $\propto \exp (-p^2/p_{\max }^2)$ (dash-dotted curve) and the sharp
cut-off (dotted curve).}
\label{Fig6}
\end{figure*}

\begin{figure*}[tbp]
\centering
\includegraphics[width=17.0cm]{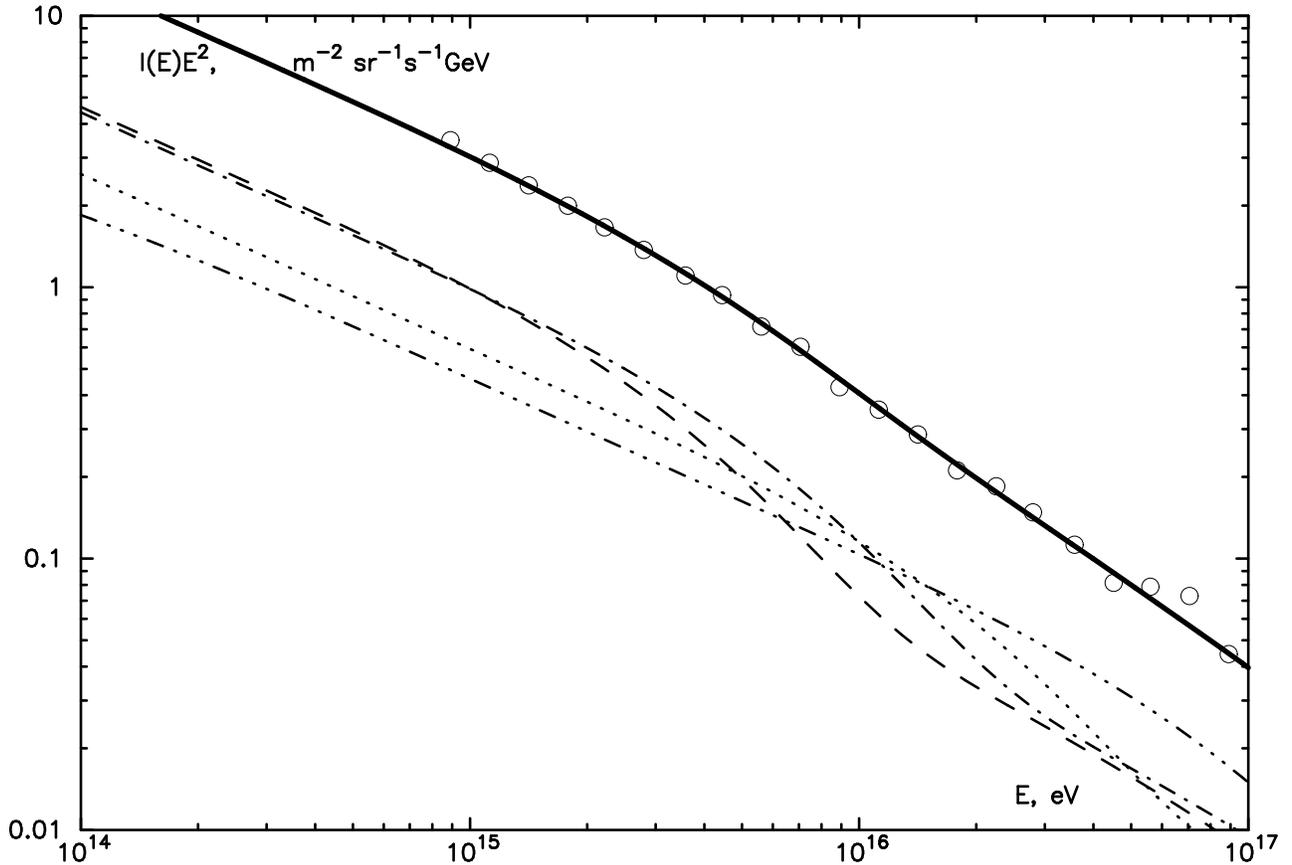}
\caption{Calculated differential spectral flux $I(E)$ (in units of m$^{-2}$%
sr $^{-1}$s$^{-1}$GeV$^{-1}$) of the CR protons (dashed curve), helium
nuclei (dash-dotted curve), carbon (dotted curve), iron (dash-dot-dotted
curve), all-particle (solid curve) in the Galaxy for the exponential
cut-off, and the all-particle spectral flux observed by the KASCADE
collaboration (empty circles). The chemical composition has been fixed at $%
E=9\cdot 10^{14}$ eV from Fig.5 of Kampert et al. (\protect\cite{kampert}).
The data for the all-particle spectrum are also taken from Kampert et al. (
\protect\cite{kampert}).}
\label{Fig7}
\end{figure*}

\section{Acceleration of the Wind-CRs}

For the sake of simplicity we shall consider CR propagation in the Galactic
Wind equatorial plane. The evolution of the isotropic part of the CR
momentum distribution function $N(r,t,p)$ inside the termination shock $r<R_{%
\mathrm{s}} $ is given by 
\[
\frac{\partial N}{\partial t}=\frac{1}{r^{2}}\frac{\partial }{\partial r}
r^{2}D_{\parallel }\cos ^{2}\alpha \frac{\partial N}{\partial r}-u\frac{
\partial N}{\partial r}+ 
\]
\[
+\frac{2up}{3r}\frac{\partial N}{\partial p}+Q(p)\frac{\delta (r-r_{0})}{
4\pi r_{0}^{2}}+ 
\]
\begin{equation}
+\frac{\Delta u}{L}\left( \frac{p}{3}\frac{\partial N}{\partial p}-\frac{1}{
\ln \sigma _{\mathrm{s}} }\int\limits_{0}^{p}\frac{dp^{\prime }}{p^{\prime }}
\left( \frac{p^{\prime }}{p}\right) ^{\gamma _{\mathrm{s}} }\frac{%
N(p^{\prime })-N(p)}{\ln (p^{\prime }/p)}\right)  \label{dN}
\end{equation}
Here $p$ is the momentum of the particle, $D_{\parallel }$ is the cosmic ray
diffusion coefficient along the magnetic field, $\alpha $ is the angle
between the magnetic field and the radial direction, $u$ is the Galactic
Wind velocity, and $Q(p)$ describes the source of CR particles at the radial
distance $r_{0}$=15 kpc from the galactic centre. The CR distribution
function $N$ is normalized as $n=4\pi \int p^{2}dpN$, where $n$ is the CR
number density. The last term on the r.h.s. of Eq. (\ref{dN}) describes the
collective acceleration by multiple shocks (see Appendix B for details). It
combines the adiabatic energy losses of the particles between the SIR shocks
of compression ratio $\sigma _{\mathrm{s}} $ with multiple reacceleration at
these shock fronts (the first and second terms in the round brackets,
respectively).

The boundary condition at the termination shock is given by the continuity
of $N$ and of the flux density:

\begin{equation}
D_{\parallel }\cos^{2}\alpha \left. \frac{\partial N}{\partial r}\right|
_{r=R_{\mathrm{s}} }+u\left(1-\frac{1}{\sigma }\right) \frac{p}{3}\left. 
\frac{ \partial N}{\partial p}\right| _{r=R_{\mathrm{s}} }=0.
\label{boundary}
\end{equation}

Note that the termination shock is a reflecting boundary in so far as the
downstream diffusion coefficient is taken to be zero in this equation.

The acceleration efficiency of the shock ensemble does not depend on energy.
Since the loss of particles into the downstream region of the termination
shock is energy independent also, we expect a power law spectrum of the
reaccelerated particles, the Wind-CRs. In order to obtain the corresponding
spectral index $\gamma $ we multiply Eq. (\ref{dN}) by the volume element $4
\pi r^{2} dr$ and integrate over $r$ from zero to $R_{\mathrm{s}} $. In the
steady state, using boundary condition (\ref{boundary})\ and assuming weak
modulation inside the termination shock as well as a power law momentum
spectrum for the Wind-CRs, we obtain for the index $\gamma $ the equation:

\begin{equation}
\gamma \left( 1+\frac{R_{\mathrm{s}} \sigma \Delta u}{3Lu}\right) +\frac{R_{%
\mathrm{s}} \sigma \Delta u}{Lu}\frac{\ln \left( 1-\gamma /\gamma _{\mathrm{%
s }} \right) }{\ln \sigma _{\mathrm{s}} } =0.  \label{gamma}
\end{equation}
The index $\gamma $ is always between 3 and the single shock spectral index $%
\gamma _{\mathrm{s}} =3 \sigma _{\mathrm{s}} /{\left( \sigma _{\mathrm{s}}
-1\right) }$. We obtained numerical solutions of Eq. (\ref{dN}) for the
Galactic Wind velocity $u$=300 km s$^{-1}$, $R_{\mathrm{s}}$=300 kpc, $%
\sigma =3$, $m=2$, $u_{\mathrm{s}} =450$ km s$^{-1}$, $\Delta u=100$ km s$%
^{-1}$, $\Omega _{\mathrm{p}}=30$ km s$^{-1}$ kpc$^{-1}$, $\Omega =20$ km s$%
^{-1}$ kpc$^{-1}$. We also take $\gamma _{\mathrm{s}} =6$ which corresponds
to the SIR shock compression ratio $\sigma _{\mathrm{s}} =2$. These numbers
determine the solution $\gamma =4.9$ of Eq. (\ref{gamma}). The wave
amplitude approximately corresponds to the numerically obtained value from
the last section. If we assume that the Galactic disk rotates with a
velocity of 200 km s$^{-1}$ , independent of Galactocentric distance, then
condition (\ref{cond}) will be satisfied for the Galactic Wind flow that
originates between 9 and 11 kpc.

The scattering of the Wind-CRs with energies larger than 10$^{16}Z$ eV on
self-generated Alfv\'{e}n waves is ineffective, since the mean free path
corresponding to the diffusion coefficient (\ref{diff}) is larger than 300
kpc. Higher energy particles can be scattered by elongated magnetic field
inhomogeneities in the Galactic Wind flow. The mean free path of these
latter particles can be of the order of length scale of these
inhomogeneities. To take this effect into account we used the diffusion
coefficient

\begin{equation}
D_{\parallel }=D_{\parallel }^{\mathrm{s}} \frac{c\lambda }{3} \left/ \left(
D_{\parallel}^{\mathrm{s}} +\frac{c\lambda }{3}\right) \right. .
\end{equation}
For numerical calculations we take the energy independent mean free path $%
\lambda =2r$. Also, a spectral index of Disk-CR sources $\gamma _{\mathrm{d}
}=4.0$ and a self-consistent cosmic ray diffusion coefficient $D_{\parallel
}^{\mathrm{s}} =10^{27}p/(m_{\mathrm{p}}c)$ cm$^{2}$s$^{-1}$ independent of $%
r$ were used. These values approximately correspond to those obtained in the
self-consistent model of CR propagation in the Galaxy (Ptuskin et al. \cite
{ptuskin}). The high energy cut-off of Disk-CRs was taken as $p_{\max
}=3\cdot 10^{6} m_{\mathrm{p}}c$.

We used the spectrum of Disk-CRs sources $Q(p)\sim p^{-\gamma _{\mathrm{d}
}}\exp (-p/p_{\max })$. The value of the maximum momentum of Disk-CRs
sources $p_{\max }$ was taken as $p_{\max }=3\cdot 10^{6} m_{\mathrm{p}}c$.

Numerical results are shown in Figs. 4 and 5. Fig. 4 shows the spectral
energy distribution (SED) of CR protons at the termination shock (dashed
curve) and inside the Galaxy (solid curve), respectively, for the case
without reacceleration. Note that the resulting acceleration at the
termination shock is weak because of the assumed large diffusion coefficient
inside the termination shock. Results taking reacceleration into account are
shown in Fig. 5. The high energy power-law tail appears beyond the Disk-CR
cut-off. The mismatch of the SED by a factor 3 inside the Galaxy is
explained by the modulation of the accelerated particles in the Galactic
Wind: the Wind-CRs must diffusively penetrate into the inner wind region
against the expanding flow.

We found that the strength of this mismatch depends on the unknown form of
the cut-off. This feature is demonstrated in Fig.6 that compares SED of
protons in the Galaxy for exponential cut-off (solid line), sharp cut-off
(dotted line) and for the cut-off $\propto \exp (-p^2/p_{\max }^2)$
(dash-dotted line).

The differential spectral flux calculated for the different cosmic ray
nuclei and the corresponding all-particle spectral flux, as well as the
experimental all-particle spectrum measured by the KASCADE collaboration
(Kampert et al. ( \cite{kampert})), are shown in Fig.7. The KASCADE
collaboration has since then presented two additional analyses (Antoni et
al. \cite{antoni}, Roth et al. \cite{roth}). For the all-particle spectrum
they agree within less than 20 percent with the data shown here. 

In principle we should use the source power of the Disk CRs as an
observational boundary condition in the Galaxy, and then calculate the
particle spectrum below the knee. Instead, we have fitted the amplitude of
the proton spectrum in Eq. (\ref{dN}) to the observations just below the
knee. This fitted proton source power of $7\cdot 10^{40}$ erg s$^{-1}$ is
within 20 $\% $ the same as that deduced by Ptuskin et al.
(\cite{ptuskin}). To put this number into perspective, we assume an 
average mechanical energy $E_{SN} = 10^{51}$~erg to be released per SN 
explosion, and a 10 percent average efficiency $\Theta = 10^{-1}$ of 
conversion of this energy into CR energy. The resulting Galactic Supernova 
rate is then $\nu_{SN} \approx 1/(45~\mathrm{yr})$.

\section{Conclusion}

This paper can be considered as an extension of our previous models for the
dynamics of the Galactic Wind and for CR propagation in the Galaxy
(Breitschwerdt et al. \cite{breitschwerdt91, breitschwerdt93}, Zirakashvili
et al. \cite{zirak}, Ptuskin et al. \cite{ptuskin}). We believe that now a
convincing physical picture for the Galactic halo, including gas, magnetic
field and CRs exists. CR sources in the Galactic disk produce energetic
particles -- the Disk-CRs -- which mainly drive the Galactic Wind flow with
its frozen-in magnetic field. CR streaming along magnetic field lines
excites Alfv\'{e}n waves. Nonlinear damping of these waves results in gas
heating up to temperatures of the order of a million degrees in the galactic
Halo (cf. Zirakashvili et al. \cite{zirak}). The balance of wave damping and
wave excitation determines the level of Alfv\'{e}nic turbulence which in
turn determines the scattering efficiency and hence the diffusion of the CR
particles. The CR spectrum observed up to the maximum particle energies
resulting from the source spectra produced by acceleration in Supernova
Remnants (presumably $E<E_{\mathrm{knee}}\sim 3\cdot 10^{15}Z$ eV) can be
fully explained in this self-consistent nonlinear model. For the explanation
of the higher energy part of the spectrum we propose a reacceleration
mechanism in the Galactic Wind flow for the most energetic particles from
the disk. The lower energy CRs are produced more effectively in the Galactic
spiral arms and therefore should drive a more powerful wind above the arms.
The interaction of these fast streams with the slow streams from the
interarm regions can result in shock formation at large distances from the
Galaxy. These shocks are similar to the Corotating Interaction Region shocks
of the outer Solar Wind. CR particles from the disk can be reaccelerated on
these shocks by a diffusive shock acceleration mechanism to generate the
Wind-CRs. The discussion of the very simple results shown in Fig. 5 and
Fig.7 allows us to conclude that the observations do not contradict the idea.

We should underline the very important role of the termination shock. The
expected strong turbulence associated with it results in an energy
independent loss of particles with energies smaller than about $10^{17}Z$\
eV. In practice this is the maximum energy for the model considered. Higher
energy particles require accelerating sources outside the termination shock.
Alternatively they may be the ubiquitous result of the decay of superheavy
particles left over from the early Universe.

\appendix

\section{Influence of the termination shock on high energy particles.}

We consider a spherically symmetric wind flow with velocity $u$. Beyond the
termination shock at radius $R_{\mathrm{s}} $ the diffusion coefficient
fulfills the inequality $D<<uR_{\mathrm{s}} $ if the turbulence level is
sufficiently high. Interior to this dissipation layer we have an interplay
between diffusion and advection. The radius of the diffusion-advection
boundary $R_{\mathrm{da}} (p)$ is determined by the condition $D(p,R_{%
\mathrm{da}} )\sim uR_{\mathrm{da}} $. Inside this zone advection is small
and will be neglected. The solution of the pure diffusion equation between $%
r_{0}$ and $R_{\mathrm{da}} $ is

\[
N(p,r)=N_{0}+ 
\]
\begin{equation}
+(N_{\mathrm{da}} -N_{0})\int\limits_{r_{0}}^{r}\frac{dr^{\prime }}{
r^{\prime }{}^{2}D(p,r^{\prime })}\left/ \int\limits_{r_{0}}^{R_{\mathrm{da}
} }\frac{ dr^{\prime }}{r^{\prime }{}^{2}D(p,r^{\prime })}\right.
\end{equation}

Here $N_{\mathrm{da}} $ and $N_{0}$ are the CR distribution functions at the
distances $R_{\mathrm{da}} $ and $r_{0}$ (Galactic disk radius),
respectively. The boundary condition at $r=r_{0}$ and the continuity
condition at $r=R_{\mathrm{da}} $ are

\begin{equation}
\left. D\frac{\partial N}{\partial r}\right| _{r=r_0}=-Q(p)
\end{equation}

and

\begin{equation}
\left. D\frac{\partial N}{\partial r}\right| _{r=R_{\mathrm{da}} } = u\frac{%
p }{3}\frac{\partial N}{\partial p}.
\end{equation}

As a result

\begin{equation}
N_{\mathrm{da}} =3\frac{r_{0}^{2}}{u}\int\limits_{p}^{\infty }\frac{
dp^{\prime }}{p^{\prime }}\frac {Q(p^{\prime })}{R_{\mathrm{da}}
^{2}(p^{\prime})},
\end{equation}

\begin{equation}
N_{0}=3\frac{r_{0}^{2}}{u}\int\limits_{p}^{\infty }\frac{ dp^{\prime }}{
p^{\prime }}\frac {Q(p^{\prime})}{R_{\mathrm{da}} ^{2}(p^{\prime})}
+r_{0}^{2}Q(p)\int\limits_{r_{0}}^{R_{\mathrm{da}} (p)}\frac{dr^{\prime }}{
r^{\prime }{}^{2}D(p,r^{\prime })}.
\end{equation}

If the diffusion coefficient is inversely proportional to $r^{2}$ (this is
the case in the self-consistent model of Ptuskin et al. (\cite{ptuskin})
because of the factor $\cos ^{2}\alpha $), the integral over $r^{\prime }$
in the last expression is determined by the upper limit and is comparable
with the first term. A source momentum spectrum with index 4.0 and a
diffusion coefficient proportional to $p$ give the value $14/3 \simeq 4.67 $
for the index of the observable spectrum .

For larger energies the radius $R_{\mathrm{da}} (p)$ is comparable with $R_{%
\mathrm{s}} $ and limited by it. Therefore, for these energies

\begin{equation}
N_{0}=3\frac{r_{0}^{2}}{uR_{\mathrm{s}} ^{2}}\int\limits_{p}^{\infty }\frac{
dp^{\prime }}{p^{\prime }}Q(p^{\prime
})+r_{0}^{2}Q(p)\int\limits_{r_{0}}^{R_{\mathrm{s}} }\frac{ dr^{\prime }}{
r^{\prime }{}^{2}D(p,r^{\prime })}
\end{equation}

This means that the momentum spectrum becomes hard like the source spectrum.

If, for example, the diffusion coefficient was independent of $r$, then the
second term can dominate the first term for $uR_{\mathrm{s}} <<D<<uR_{%
\mathrm{s}} ^{2}/r_{0}$ in the last equation. But in this case the spectrum
inside the Galaxy would be determined by diffusion only and one should take
for example a value of 4.0 for the source index and the diffusion
coefficient proportional to $p^{0.67}$ in order to reproduce the observable
spectrum. It becomes again hard for high enough energies $D>>uR_{\mathrm{s}}
^{2}/r_{0}$.

\section{CR acceleration by the SIR shock system}

Let us investigate the diffusive acceleration by a periodic shock wave. Let
the velocity profile be a continuous function $u(x)$ for $0<x<L$ where $L$
is period of the wave. Shocks are located at $x=0$, $x=L$, $x=2L$, etc. At
each shock the flow velocity drops from the value $u_{1}=u(L-0)$ to the
value $u_{2}=u(0+0)$.

We shall consider the case of a small diffusion coefficient, characterized
by the inequality $D<<uL$. In this case the particle is convected by the
flow from one shock to the neighboring one, where it is diffusively
accelerated. Subsequently it is convected to the next shock where it is
accelerated again, etc.

Let $N(p)$ be the particle distribution function in momentum $p$. It is
normalized as $\int 4\pi p^{2}dpN(p)=n$, where $n$ is the particle number
density. Near the shock particles are accelerated and $N(p)$ changes
correspondingly. The relation between upstream \ and downstream
distributions is given by diffusive acceleration theory (e.g. Blandford \&
Ostriker \cite{blandford})

\begin{equation}
N_{\mathrm{down}}(p)=\gamma _{\mathrm{s}} \int\limits_{0}^{p}\frac{%
dp^{\prime }}{p^{\prime }} \left( \frac{p^{\prime }}{p}\right) ^{\gamma _{%
\mathrm{s}} }N_{\mathrm{up}}(p^{\prime }),
\end{equation}
where $\gamma _{\mathrm{s}} =3\sigma _{\mathrm{s}} /(\sigma _{\mathrm{s}}
-1) $ is the power law index of particles accelerated by a single shock with
compression ratio $\sigma _{\mathrm{s}} =u_{1}/u_{2}$. For our purposes it
is more convenient to use the Mellin transform $F(s)=\int\limits_{0}^{\infty
}dpp^{s-1}N(p)$. Then $F_{\mathrm{down}}(s)=F_{\mathrm{up}}(s)/(1-s/\gamma _{%
\mathrm{s}} )$. Between the shocks a particle loses energy adiabatically.
The distribution function $N(p)$ then transforms into $N(p\sigma _{\mathrm{s}
} ^{1/3})$. For the Mellin transform such a transformation reduces to a
multiplication with $\sigma _{\mathrm{s}} ^{-s/3}$. Hence, after $n$
acceleration cycles the Mellin transform of the distribution function is
given by

\[
F_{n}(s)=F_{0}(s)\sigma _{\mathrm{s}} ^{-sn/3}/(1-s/\gamma _{\mathrm{s}}
)^{n}= 
\]
\begin{equation}
=F_{0}(s)\exp \left( -n\left( \frac{s}{3}\ln \sigma _{\mathrm{s}} +\ln
\left( 1-s/\gamma _{\mathrm{s}} \right) \right) \right)
\end{equation}

One can check that this does not depend on the initial position of the
particle. It can be near the shock or somewhere between the shocks. Let us
now introduce the cycle duration $T=\int\limits_{0}^{L}dx/u(x)$. Then, at
the time $t=nT$, the function $F_{n}(s)$ can be found as the solution of the
equation

\begin{equation}
\frac{\partial F(s)}{\partial t}=-\frac{F(s)}{T} \left( \frac{s}{3}\ln
\sigma _{\mathrm{s}}+\ln \left( 1-s/\gamma _{\mathrm{s}}\right) \right)
\label{mresult}
\end{equation}
with the initial condition $\left. F(s)\right| _{t=0}=F_{0}(s)$. This
equation for the Mellin transform $F(s)$\ corresponds to the following
equation for the distribution function $N(p)$: 
\begin{equation}
\frac{\partial N}{\partial t}=\frac{1}{T}\left[ \frac{p}{3}\frac{\partial N}{
\partial p}\ln \sigma _{\mathrm{s}}-\int\limits_{0}^{p} \frac{dp^{\prime }}{
p^{\prime } }\left( \frac{p^{\prime }}{p}\right) ^{\gamma _{\mathrm{s}}} 
\frac{N(p^{\prime })-N(p) }{\ln (p^{\prime }/p)}\right]  \label{result}
\end{equation}
It is simple to verify that the Mellin transform of this equation gives Eq.
( \ref{mresult}). In this procedure the value of integral 
\[
\int\limits_{0}^{\infty }\frac{dy}{y}\left( e^{-ay}-e^{-by} \right) =\ln
(b/a),\quad a>0,b>0 
\]
should be used.

It is important that the adiabatic energy losses described by the first term
in square brackets of Eq. (\ref{result}) enter this equation additively.
This derivation is also valid in the presence of energy independent
additional energy losses or escape of particles. The corresponding terms
should be added to the right-hand side of Eq. (\ref{result}).

For weak shocks $\sigma _{\mathrm{s}}-1<<1$\ the operator on the r.h.s. of
Eq. (\ref{result}) is reduced to diffusion in momentum space. The
corresponding diffusion coefficient is given by

\begin{equation}
D_{pp}=\frac{p^{2}}{18}\frac{(\Delta u)^{2}}{v_{\mathrm{s}}L},
\end{equation}
where $v_{\mathrm{s}}$ is the shock velocity and $\Delta u=u_{1}-u_{2}$ is
the velocity jump at the shock.

In the case of a saw-tooth wave velocity profile

\begin{equation}
u(x)=u_{2}+(u_{1}-u_{2})\frac{x}{L}
\end{equation}
$T=L\ln \sigma _{\mathrm{s}}/\Delta u$, and the r.h.s. of Eq. (\ref{result})
reduces to the last term in Eq. (\ref{dN}) of the main text.

\begin{acknowledgements}
The work of VNZ was done during his visit at the Max-Planck-Institut
f\"ur Kernphysik in Heidelberg under the auspices of the
Sonderforschungsbereich 328 and was also supported by the Russian
Foundation of Basic Researches grant 01-02-17460. We thank
an anonymous referee for many important comments.

\end{acknowledgements}

\end{document}